\documentclass[fleqn,usenatbib]{mnras}

\usepackage{newtxtext,newtxmath}
\usepackage[T1]{fontenc}
\usepackage{ae,aecompl}

\usepackage{graphicx}	% Including figure files
\usepackage{amsmath}	% Advanced maths commands
\usepackage{amssymb}	% Extra maths symbols

\newcommand\eqncomma{\, ,}
\newcommand\eqnstop{\, .}

\newcommand\invhmpc{h^{-1}\textrm{Mpc}}

\title[Finding galaxy groups using Markov Clustering]{A new approach to finding galaxy groups using Markov Clustering}

\author[L. Stothert et al.]{
L. Stothert$^{1,2}$,
P. Norberg$^{1,2}$,
C. M. Baugh$^{1}$.
\\
$^{1}$Institute for Computational Cosmology, Department of Physics, Durham University, South Road, Durham DH1 3LE, UK \\
$^{2}$Centre for Extragalactic Astronomy, Department of Physics, Durham University, South Road, Durham DH1 3LE, UK \\
}

\date{Accepted XXX. Received YYY; in original form ZZZ}

\pubyear{2018}

\begin{document}
\label{firstpage}
\pagerange{\pageref{firstpage}--\pageref{lastpage}}
\maketitle

% Abstract of the paper
\begin{abstract}
We present a proof of concept of a new galaxy group finder method, Markov graph CLustering \citep[MCL;][]{mcl} that naturally handles  probabilistic linking criteria. We introduce a new figure of merit, the variation of information statistic \citep[VI;][]{melia2003}, used to optimise the free parameter(s) of the MCL algorithm. We explain that the common Friends-of-Friends (FoF) method is a subset of MCL. We test MCL in real space on a realistic mock galaxy catalogue constructed from a N-body simulation using the \texttt{GALFORM} model. With a fixed linking length FoF produces the best group catalogues as quantified by the VI statistic. By making the linking length sensitive to the local galaxy density, the quality of the FoF and MCL group catalogues improve significantly, with MCL being preferred over FoF due to a smaller VI value. The MCL group catalogue recovers accurately 
%(to within 7\%) 
the underlying halo multiplicity function at all multiplicities. MCL provides better and more consistent group purity and halo completeness values at all multiplicities than FoF.
As MCL allows for probabilistic pairwise connections, it is a promising algorithm to  find galaxy groups in photometric surveys.
\end{abstract}

% Select between one and six entries from the list of approved keywords.
% Don't make up new ones.

%%%%%%%%%%%%%%%%%%%%%%%%%%%%%%%%%%%%%%%%%%%%%%%%%%

%%%%%%%%%%%%%%%%% BODY OF PAPER %%%%%%%%%%%%%%%%%%

\section{Introduction}
The fundamental assumption behind galaxy formation theory is that galaxies form inside dark matter haloes \citep{White1978}. The hierarchical assembly of haloes and the timescale for galaxy mergers means that halos often have a main or central galaxy, accompanied by distinct satellite galaxies. There are clear predictions for the properties of the galactic content of halos that can be tested if we can identify a high fidelity sample of galaxy groups from galaxy surveys that retains a connection to the underlying dark matter halos \citep{eke2004,Eke2005,vdb2005,yang2005}. 

The identification of a galaxy group requires an algorithm to associate galaxies with a common, unique dark matter halo. Many ways have been explored to do this, with the most common being Friends-of-Friends \citep[FoF\,; e.g. ][]{huchra82,zeldovich82}. For example, \cite{eke2004} and \cite{robothamgroups} created FoF galaxy group catalogues from the 2dF Galaxy Redshift Survey \citep{2dfgrs} and the Galaxy And Mass Assembly survey (GAMA)  \citep{gamadriver}. \cite{pFoF} extended FoF for galaxies with photometric redshifts, which was then applied to the Pan-STARRS1 medium deep survey \citep{pFoFpanstaars}. \cite{yanggroupfinder} developed a halo based group finder that was used to construct a group catalogue using Sloan Digital Sky Survey (SDSS) galaxies \citep{sdssgroups}.

However, despite the success of FoF-based methods they are far from perfect and struggle when applied to low density samples as is the case with galaxy catalogues. This should be contrasted with their application  to numerical simulations where the particle distribution is thousands of times denser (if not more) than a typical galaxy distribution. When applied to galaxy catalogues, FoF tends to create either too many low multiplicity groups (by fragmentation of the larger ones) or groups that are too big (by spuriously joining smaller groups to bigger ones). Measures of purity and completeness are then used to rate the quality of the group catalogue and these statistics tend to be combined in some way, to create a statistic that should be minimized to ensure an `optimal' set of groups (see, for example, \citealt{eke2004}). It is worth noting that FoF does not use all of the available pairwise information, nor can it be extended naturally to handle probabilistic positional information, as is the case with e.g. photometric redshifts.

Here we show that the FoF approach to galaxy group finding is just one solution to the graph clustering problem \citep[e.g.][]{graphclustering}. Graph clustering aims to find clusters of points given all pairwise connection amplitudes between them. It is a problem that occurs in many situations, such as detecting communities in social networks \citep[e.g.][]{socialnetworkclustering}. We explain, in Section~\ref{sec:mcl}, how the FoF algorithm is a subset of the Markov graph CLustering algorithm MCL \citep{mcl}, which we apply to the problem of galaxy group detection. MCL has been widely used in the field of bioinformatics in detecting groups of proteins based on their pairwise interactions \citep[e.g.][]{markovclusteringbiology}.

Our overall aim is to construct a group catalogue using the narrow band PAU Survey \citep[PAUS; e.g.][]{eriksen2018,stothert2018}. A PAUS group catalogue would probe significantly fainter galaxies than one built using SDSS or GAMA, and would cover a larger area with better completeness in both sampling and redshift than a group catalogue constructed using similar depth surveys such as zCOSMOS \citep{zcosmos} or VIPERS \citep{vipers}. Hence a PAUS group catalogue would provide a better probe of the redshift evolution of halos as traced by galaxy groups and better sampling of low mass halos. The challenge with finding galaxy groups in PAUS lies in the varying accuracy of the PAUS photometric redshifts. MCL is a promising approach as it allows probabilistic pairwise connections \citep[see also][for another approach]{tempel2018}, something that could be useful for PAUS where it is more natural to frame pairwise connections as probabilities than as binary links.

%{PN: Not sure whether this is best here or in the conclusions -- CMB doesn't fit in here}
%While working on Markov Clustering, \cite{tempel2018} proposed a new Bayesian group finder based on marked point processes. Both approaches are %vastly different in nature, but might have similar positive prospects in the sense that they are set up to use probabilistic spatial information.

Section~\ref{sec:mcl} presents the MCL algorithm and explains its relation to the standard FoF algorithm. Section~\ref{sec:mock} presents the mock catalogue which is used to test the algorithm. Section~\ref{sec:stats} summarises the metrics we use to assess the group finding performance. Section~\ref{sec:tests} presents the results in real space. We provide our conclusions and future prospects of the Markov CLustering algorithm MCL in Section~\ref{sec:conclusion}. Hereafter we refer to a `clustering' of galaxies interchangeably with a `grouping' of galaxies. 
%We refer to the two point correlation function when discussing galaxy clustering in the more common context. 
Throughout we assume a flat $\Lambda$CDM cosmology, with parameters $\Omega_{\rm m}=0.272$, $\sigma_8=0.81$ and $h=0.704$, consistent with those used to create the mocks (as described in Section~\ref{sec:mock}). We refer the reader to \cite{stothert2018b} for additional details regarding the algorithm, the mocks and some of the additional tests performed (and not reported here).

\section{Markov Clustering}
\label{sec:mcl}

The Markov CLustering algorithm (MCL) was developed as a fast, scalable approach to graph clustering\footnote{The MCL code is publicly available at \url{http://micans.org/mcl/}.} \citep{mcl}. Graph clustering~\citep[e.g.][]{graphclustering} is a solution to the problem of finding clusters of points given their pairwise connection amplitudes. One obvious and instructive example of a graph clustering problem is detecting communities within a social network~\citep{socialnetworkclustering}. Here users are `friends' with other users. The entire friendship network can be represented by a (symmetric) binary matrix, which we call the pairwise connection matrix $W$, with elements $w_{ij}$. If users $i$ and $j$ are friends, $w_{ij}$ is 1 and is 0 otherwise. A graph clustering algorithm detects communities within this structure. MCL was chosen for two key reasons: (1) in one of its limits it tends to the standard FoF algorithm as explained later; (2) it supports probabilistic pairwise connections rather than just fixed binary links, which is essential for finding galaxy groups with photometric redshifts.

The MCL algorithm has one free parameter, the inflation parameter $\Gamma$, which has to be greater than or equal to unity. The algorithm takes the initial pairwise connection matrix,  $W_0$ (specified by its elements $w_{ij}^0$), as an input and assigns points to clusters following an iterative process, where $W_k$ is the pairwise connection matrix after $k$ steps:

\begin{enumerate}
\item{Normalise $w_{ij}^0$ column-wise such that \ $\sum_j w_{ij}^0=1$.} 
\item{At step $k$, create $W_k$ by squaring the pairwise connection matrix $W_{k-1}$, i.e.\ $W_k=W_{k-1}^2$.}
\item{Raise every element of $w_{ij}^k$ to the power of $\Gamma$, i.e.\ $(w_{ij}^k)^{\Gamma}$}
\item{Renormalise $w_{ij}^k$ column-wise such that \ $\sum_j w_{ij}^k=1$.}
\item{Repeat from (ii) until all elements of $W_k$ have  converged individually to within a specified tolerance.} 
\item{Rearrange the converged cleaned $W_k$ matrix into a block diagonal matrix and read off the groups.}
\end{enumerate}

\noindent
We now explain each step in turn. The initial column-wise normalisation in step (i) above -- and those that follow in step (iv) -- are necessary to ensure that the pairwise connection elements relating to point $i$ can be treated as probabilities. 
%{Note that for a symmetric matrix (as is most often the case) a column-wise normalisation is equivalent to a row-wise normalisation}. 
By squaring the pairwise connection matrix $W_{k-1}$ to create a new pairwise connection matrix, $W_k$, the MCL algorithm approximately simulates a random walk on the graph by using the elements $w_{ij}^k$ as transition probabilities to determine which pairs are more bound than others.\footnote{See e.g.\ \citet{mcl} for a discussion of why this approach produces a similar result to a standard random walk, while strictly speaking it is not a random walk.} 
%{It does this by simulating a random walk on the graph using $w_{ij}$ as transition probabilities to determine which points are most bound. A random walk will get temporarily stuck, more so in a structure that is tightly bound, only rarely jumping between structures. This is not strictly a random walk as the initial matrix is not saved. See \cite{mcl} for a discussion on why this produces a similar result and why this exact procedure was chosen}.
%
Step (iii), raising the elements of $w_{ij}^k$ to the power  $\Gamma$, is designed to boost the more travelled connections and reduce the less travelled inter-cluster ones. %This effect is to some extent guaranteed through the column-wise normalisation applied in step (iv). 
This process of matrix multiplication (here assumed to be squaring), element inflation (to the power of $\Gamma$) and column-wise normalisation is repeated until a predefined convergence criteria is met by the pairwise connection matrix $W_k$. 
The convergence criterion is that the final matrix becomes idempotent, i.e.\ invariant under expansion and inflation. The exact criterion is expressed in terms of the maximum over all columns of the difference between the maximum value in a column and the sum of all elements squared of that column.
Once converged, the matrix $W_k$ is cleaned (by setting to zero all $w_{ij}^k$ elements below a pruning value of 10$^{-4}$) and then rearranged with row replacement into a block diagonal matrix, with members of each group defined by the matrix blocks.

\begin{figure}
  \includegraphics[width=\linewidth,trim=0 5 30 30,clip]{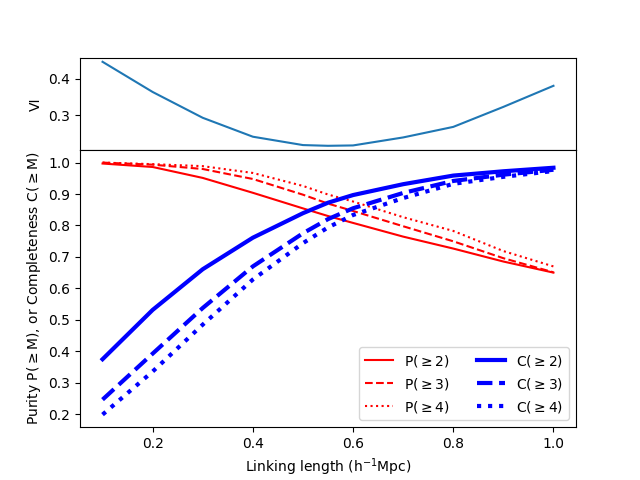}
  \caption{Variation of information (VI; top), group purity and halo completeness (P($\geq$M) and C($\geq$M); main  panel) as a function of linking length in a standard FoF approach to galaxy group finding for three values of minimum group/halo multiplicity M. The minimum of VI provides a good compromise between group purity and halo completeness at all multiplicities.
  %The best fit value of linking length relative to the mean galaxy separation found in \cite{eke2004} would give a linking length of 0.6 $\invhmpc$ in this catalogue, which roughly agrees with our minimum variation of information value.
  }
  \label{fig:vi_pure_FoF}
\end{figure}

\begin{figure*}
  \includegraphics[width=0.9\linewidth,trim=0 18 0 10 50,clip]{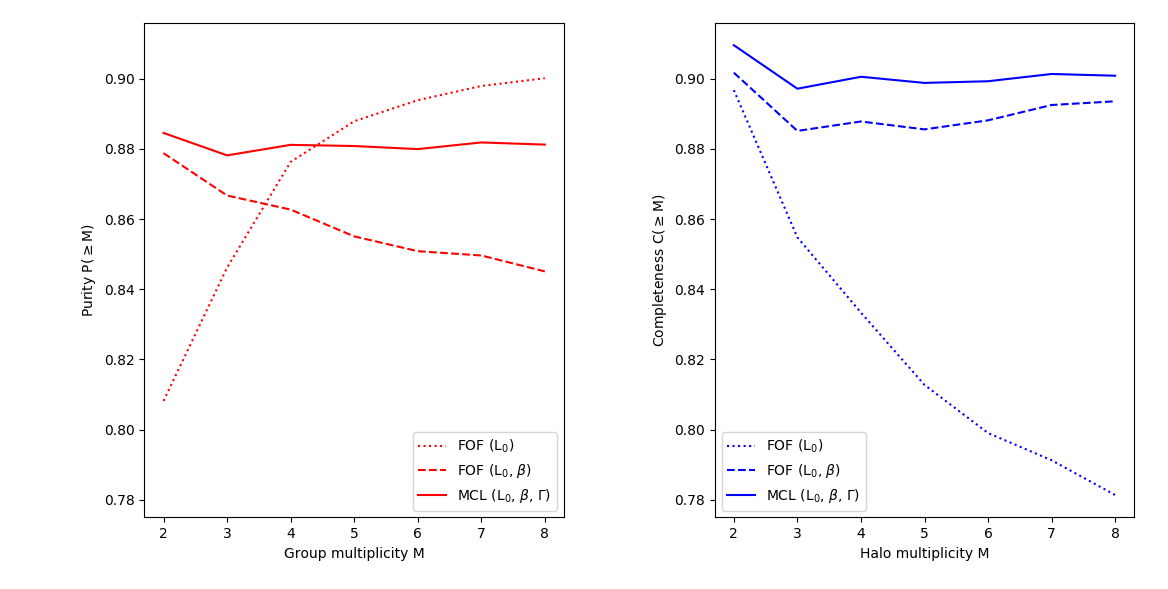}
  \caption{Group purity, P($\geq$M) (left panel), and halo completeness, C($\geq$M) (right panel), as a function of minimum multiplicity, M, for different VI minimised galaxy group catalogues: simple FoF (dotted), FoF with density enhancement (dashed) and MCL with density enhancement group catalogues (solid). The purity and completeness of the MCL group catalogue is the most consistent as a function of multiplicity, and has undoubtedly the best halo completeness.}
  \label{fig:purity_completeness_size}
\end{figure*}

%Some notes about MCL
At face value MCL is an iterative $N^2$ process as all links between $N$ points need to be defined at each iteration.
The larger the value of the inflation parameter, $\Gamma$, the more rapidly the pairwise connections tends towards zero during the iterations and the faster the MCL algorithm will split structures into smaller components. 
%By its nature, once a structure is split by the MCL algorithm, it remains split at all future iterations. Likewise once a structure is split by an inflation parameter $\Gamma_1$, it will always be split by any $\Gamma>\Gamma_1$.
A structure that is split by inflation parameter $\Gamma_1$ will always be split by any $\Gamma>\Gamma_1$. In principle $\Gamma$ has no maximum value but there will be a value of $\Gamma$ above which the catalogue stops splitting, as all clusters become fully connected sub-graphs with equal pairwise connections, i.e.\ all points in every cluster are connected only to all other points within the same cluster with the same $w_{ij}$ value (and such clusters are not split by MCL). We note that a $\Gamma$ value of unity will connect any structure that has any path connecting it. In that case MCL tends to converge extremely slowly as no links are ever trimmed from the matrix (see Section~\ref{sec:stats} for a practical application).
%which is used to trim connections that are used for these rare inter-cluster jumps. MCL is an iterative process that repeats matrix operations on $w_{ij}$ until the matrix converges and the clusters can be read off.

%{\bf FoF within MCL approach}
In the astrophysical case we first have a connection criterion that sets the values of $w_{ij}$ between galaxies $i$ and $j$. This is normally based on a distance criterion between two galaxies, setting $w_{ij}$ to 1 if the galaxies are closer to each other than some specified linking length and 0 otherwise. The standard FoF algorithm connects all points that could be reached via a succession of links between points. This outcome is exactly the same as that for MCL with the inflation parameter $\Gamma$ set to unity.
%\footnote{We made the test and found the results of FoF and MCL with $\Gamma=1$ to be identical. A too relaxed convergence criteria can result in MCL not probing all FoF links, typically the most tenuous ones, i.e.\ those links that are the least likely to be traversed .}
Therefore the FoF algorithm should be considered as the limit towards which MCL converges when $\Gamma$ tends to unity, i.e.\ formally FoF is a subset of MCL. An advantage of MCL over FoF is that, even though MCL like FoF uses all pairwise links, MCL gives higher priority to points that are more connected than those with fewer connections, unlike FoF. By carefully using the inflation parameter, the less well connected points (or less important pairwise links) can be broken up. Only through detailed tests on mocks (see Section~\ref{sec:tests}) can the accuracy of the MCL algorithm be assessed against e.g.\ FoF. 

\section{Mock catalogue}
\label{sec:mock}

To test the MCL approach to galaxy group finding we apply it to a realistic real space galaxy mock catalogue. We use real space rather than redshift space to better understand the impact of changing the clustering algorithm. We use a $z = 0$ snapshot of the \texttt{GALFORM} model presented in \cite{g17}, implemented in the  125 $\invhmpc$ per side MilliGas simulation cube. Note that this simulation has the same cosmology and number of snapshots as the 500$\invhmpc$ MR7 simulation \citep{guo2013}. We use a smaller simulation to speed up the calculations, as deciding between methods of linking galaxies and optimisation of free parameters requires running the algorithm many times. The catalogue is limited in the rest frame r-band to $M_r - 5\log h < -20.0$ and contains $\sim 20,000$ galaxies, corresponding to a galaxy density of $\sim 10^{-2}$ $(\invhmpc)^{-3}$, comparable to the GAMA survey at $z \sim 0.15$ ~\citep{gamadriver,liske2015,baldry2018} . By construction each galaxy belongs to a unique dark matter halo and each halo contains one or more galaxies. See \cite{stothert2018b} for further details of how the mock catalogue was constructed. 

\section{Goodness of fit metrics}
\label{sec:stats}

We assess the quality of group finding using the measures of purity and completeness. 
%Halos provide the true clustering of the simulation and groups the clustering found by the algorithm (e.g.\ FoF or MCL). 
Group purity, $P$, quantifies the extent to which galaxies in the same group are actually in the same halo \citep[e.g.][]{manning2008,wu2009}:
\begin{equation}
\label{eq:purity}
P = \frac{1}{\sum_{i = 1}^{N_G}\sum_{j = 1}^{N_H} n_{ij}} \sum_{i = 1}^{N_G} \textrm{max}_j(n_{ij}) %\eqnstop
\end{equation}
\noindent where $n_{ij}$ is the number of galaxies in group $i$ {\emph{and}} halo $j$, $N_G$ is the total number of groups and $N_H$ is the total number of halos. Similarly, we define the halo completeness, $C$, which quantifies the extent to which galaxies in the same halo are placed in the same galaxy group:
\begin{equation}
\label{eq:completeness}
C = \frac{1}{\sum_{i = 1}^{N_G} \sum_{j = 1}^{N_H} n_{ij}} \sum_{j = 1}^{N_H} \textrm{max}_i(n_{ij}) \eqnstop
\end{equation}

\noindent We also use the associated cumulative measures $C(\geq M)$ and $P(\geq M)$ defined, respectively, as the completeness of halos and the purity of groups with multiplicity (i.e.\ number of members) greater than or equal to $M$.  
For the cumulative measures, the multiplicity cut is only applied to the halos for $C(\geq M)$ and groups for $P(\geq M)$.

To optimise the parameters of the MCL algorithm a single statistic is desirable. Here we would like a problem agnostic measure to build an `optimal' group catalogue. Most astrophysical applications invoke combinations of bijective measures of completeness and purity  \citep{gerke2005,robothamgroups,knobel2012,pFoFpanstaars}. Instead we follow \cite{wu2009} who tested multiple goodness of fit metrics in a statistical context and choose to use the variation of information (VI) \citep{melia2003}. 
% To our knowledge this is the first time this metric has been used in an astrophysical context.

VI, also called the shared information distance, quantifies the distance between two clusterings by looking at the amount of information in each that cannot be inferred using the other clustering. A smaller value of VI means a better clustering, so we minimise this metric to determine the best MCL parameters. Using a definition of entropy from statistical physics, VI is formally written as
\begin{equation}
\begin{split}
\label{eq:vi}
VI = &- \sum_{j=1}^{N_H} p_{\Sigma j} \ln ( p_{\Sigma j} ) -\sum_{i=1}^{N_G} p_{i \Sigma} \ln ( p_{i \Sigma} )  \\
& - 2\sum_{i=1}^{N_G}\sum_{j=1}^{N_H} p_{ij} \ln \Bigg( \frac{p_{ij}}{p_{i \Sigma}p_{\Sigma j}} \Bigg),  %\eqnstop
\end{split}
\end{equation}
where $p_{xy}$=$n_{xy}/n_{\Sigma \Sigma}$ for any $x$ or $y$. This includes the special case of $x$=$\Sigma$ (or $y$=$\Sigma$ or $x$=$y$=$\Sigma$) for which we define $n_{\Sigma j} = \sum_{i = 1}^{N_G} n_{ij}$, $n_{i \Sigma} = \sum_{j = 1}^{N_H} n_{ij}$ and $n_{\Sigma \Sigma} = \sum_{i = 1}^{N_G} \sum_{j = 1}^{N_H} n_{ij}$, corresponding to the number of galaxies in group $j$, the number of galaxies in halo $i$ and the total number of galaxies respectively. 

We validate the use of VI by testing how it relates to the more familiar measures of halo completeness and group purity (Eqns.~\ref{eq:completeness} and ~\ref{eq:purity}). Fig.~\ref{fig:vi_pure_FoF} shows the VI and three values of $P(\geq M)$ and $C(\geq M)$ as a function of the assumed fixed linking length for a standard FoF algorithm applied to our mock galaxy catalogue. The minimum value of VI gives a catalogue that is well balanced between completeness and purity. The minimum value of VI also agrees with the value of the linking length relative to the mean galaxy separation found in e.g.\ \cite{eke2004}. This shows that our choice of optimisation statistic is sensible, and that using it in standard FoF produces results comparable to those found in previous work.

\section{Results}
\label{sec:tests}

We compare the results of applying two different clustering methods (MCL and FoF) to the mock galaxy catalogue. In each case the free parameters are found by minimizing VI (Eq.~\ref{eq:vi}). All models set the binary connection between galaxies $i$ and $j$, $w_{ij}$, to unity if the pairwise separation $r_{ij}$ is smaller than the linking length $L_{ij}$, and 0 otherwise.

\begin{figure}
  \includegraphics[width=\linewidth,trim=0 5 30 30,clip]{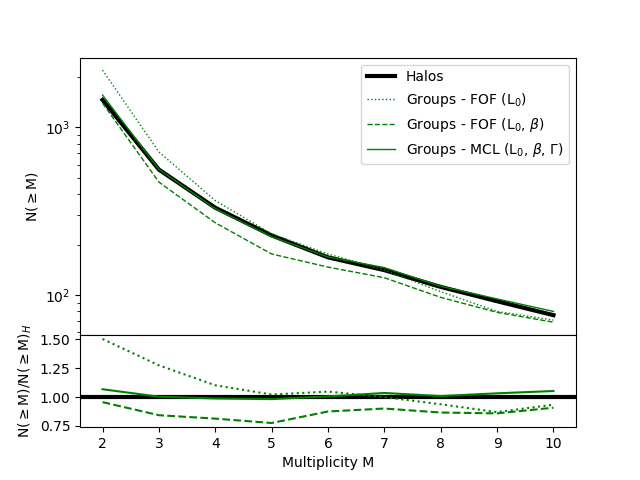}
  \caption{The cumulative multiplicity function, $N(\geq M)$, for halos (truth, thick black line) and three groups catalogues (green lines), which are the simple FoF (dotted), FoF with density enhancement (dashed) and MCL with density enhancement (solid). The bottom panel shows the ratio of the multiplicity functions to the truth, $N(\geq M)_H$. MCL with density enhancement recovers the true halo multiplicity function extremely well (to better than 7\% at all multiplicities).}
  \label{fig:final_multiplicity}
\end{figure}

%\begin{equation}
%\label{eq:connection_binary}
%    w_{ij}= 
%\begin{cases}
%    1,& \text{if } r_{ij} \leq L_{ij}\\
%    0,              & \text{otherwise} \eqnstop
%\end{cases}
%\end{equation}
%
%\noindent

In our first groupings we adopt a constant linking length, i.e.\ $L_{ij}$ is fixed. For FoF this is the only free parameter. Fig.~\ref{fig:vi_pure_FoF} indicates that the optimal value is $L_{ij} =0.55 \invhmpc$. The optimal solution with MCL using a fixed linking length is achieved, according to VI, when the inflation, $\Gamma$, tends to unity, indicating that the FoF algorithm is preferred over MCL in this fixed linking length scenario. This is because with a fixed linking length small structures have poor purity and large structures have poor completeness, and increasing $\Gamma$ only splits the larger  structures further, worsening the situation. Hence, hereafter we only show the FoF results for fixed linking length.

The second set of models use a variable linking length set by the geometric mean of the local galaxy density in an attempt to include the known scale dependence of the clustering, as was done in e.g.\ \cite{eke2004} and \cite{robothamgroups}. We calculate the local density, $\rho_i$, at the position of galaxy $i$ using a 3D Gaussian kernel with $\sigma = 1 \invhmpc$ truncated at 4$\sigma$. Other reasonable values of the smoothing scale were tested with no significant improvement found. $L_{ij}$ is now given by
\begin{equation}
\label{eq:density_enhancement_deterministic}
L_{ij} = L_0 \Bigg( \frac{\sqrt{\rho_i \rho_j}}{\left< \rho \right>(r_{ij}) } \Bigg) ^{\beta} . %\eqncomma
\end{equation}

\noindent
$L_0$ and $\beta$ are free parameters and <$\rho$>($r$) is the mean value of the geometric mean of the pairwise local densities at separation $r$ 
\begin{equation}
\label{eq:geometric_mean_def}
\left< \rho \right>(r) \equiv \frac{\sum_i\sum_j \sqrt{\rho_i\rho_j}}{\sum_i\sum_j} \eqncomma
\end{equation}

\noindent
where the sums are over all galaxies separated by $r$. This process extends the linking length for galaxy pairs in overdense region relative to those in underdense ones. A scale dependent normalisation is necessary because, for pairs of galaxies at small separations, the product of their local galaxy densities will on average be larger than that of galaxy pairs at larger separations.

The first density enhanced model connects groups using the FoF algorithm and has two free parameters, $\beta$ and $L_0$. %This model recovers the constant linking length model if $\beta = 0$. 
The best value of VI is at $\beta$ = 0.6 and $L_0$ = 0.9 $\invhmpc$. From its VI value, this best FoF density enhanced model is preferred over the best model with a constant linking length. 
%PN commented out the next sentence as not clear:
%The unmodified value of the linking length is now larger than for the constant case. This is because this scheme can avoid incorrectly connecting too many pairs of galaxies that lie in underdense regions that shouldn't be connected.

The second density enhanced model uses MCL, so adds the inflation $\Gamma$ as a third free parameter. 
%It recovers the FoF model with density enhancement if $\Gamma = 1$. 
The minimum value of VI is now given by $\Gamma = 1.6$, $\beta$ = 0.6 and $L_0$ = 1.1 $\invhmpc$. From its VI value, this optimal MCL density enhanced algorithm produces the best catalogue of the four algorithms considered (FoF and MCL, with and without density enhanced linking lengths).
%PN commented out the next sentence as not clear (same as before):
%The best model here also has a larger unmodified linking length than the comparable FoF model. The MCL algorithm allows more pairwise connections to be made, and then the poorly connected structures are split apart by the inflation parameter.

Fig.~\ref{fig:purity_completeness_size} shows the group purity $P(\geq M)$ and halo completeness $C(\geq M)$ as functions of group and halo multiplicity respectively for the optimal catalogue produced by each of the three models. FoF has low purity for small groups
%P($\geq$2) $\sim$ 0.8
and poor completeness for large halos.
%C($\geq$2) $\sim$ 0.8. 
FoF with density enhancement performs significantly better, but still tends to over-join some larger groups, explaining the fall in purity with increasing  multiplicity. The density enhanced MCL algorithm improves on both aspects and produces a group purity and halo completeness that are largely independent of multiplicity. A catalogue with high purity and completeness that are only mildly dependent on multiplicity is preferable. This MCL also produces a catalogue that has higher halo completeness for all multiplicities considered here than the corresponding FoF algorithm with density enhancement. We note that the purity of high multiplicity groups is larger for the simple FoF case, but this is at the expense of a very poor halo completeness.

Fig.~\ref{fig:final_multiplicity} shows the cumulative multiplicity function, $N(\geq M)$, for the underlying halos and the three galaxy group catalogues. By including density enhancement, FoF provides a better estimate of the number of small groups, but the number of large groups remains underestimated. MCL with density enhancement impressively recovers the correct numbers of groups at all multiplicities tested here to better than 7\%, and often to better than 3\%. This is to be compared to the best FoF performance which underestimates the number of halos by as much as 25\% from the truth at $N \geq 5$ and $\sim$15\% for most multiplicities. Note these results were not used to identify the optimal group finder, which is determined by minimizing the variation of information (VI) for each clustering model.

Our results show that MCL can better address the stochastic nature of `bridges' connecting structure that appear with FoF. FoF needs to be more cautious about the connection criterion as there is a large penalty if even a single link is found between two large structures, whereas MCL reduces this penalty by using inflation to break loosely connected structures. These `bridges' cause the number of high multiplicity FoF groups to be underestimated (see Fig.~\ref{fig:final_multiplicity}), and their group purity to be low (see Fig.~\ref{fig:purity_completeness_size}). Both aspects are improved significantly upon using MCL.

\section{Conclusions}
\label{sec:conclusion}

For the first time in an astronomical context we apply the Markov CLustering algorithm \citep[MCL;][]{mcl}, which is part of the more general graph clustering algorithms, to identify galaxy groups. MCL has one free parameter, inflation, $\Gamma$. We show that the widely used FoF algorithm is a subset of  MCL; with $\Gamma=1$, MCL produces the same result as the deterministic FoF algorithm. We apply MCL to detect galaxy groups in a real space galaxy mock catalogue. We minimize the variation of information \citep[VI;][]{melia2003} to compare group catalogues to real halos. We validate this choice by showing that the minimum value of VI for a simple FoF approach is found at linking lengths that are in good agreement with previous values \citep[e.g.][]{eke2004}. 

For a constant linking length FoF produces the best group catalogue. Nevertheless, FoF returns too many spurious small groups and too few large groups: increasing inflation away from unity only makes this discrepancy worse.
%We vary the linking length as a function of the local density of the two galaxies in a pair to address the multiplicity dependency of the results. This local density enhancement is normalised in such a way that can be measured from the real data and requires no free parameter. This scheme significantly improves the results of both the FoF and MCL approaches. Using this scheme the MCL algorithm produces the catalogue with the minimum value of VI.
Using a linking length sensitive to the local density to account for the scale dependence of the grouping, MCL is superior to FoF (i.e.\ VI is minimised with $\Gamma>1$). In both cases the group purity and halo completeness are  improved over a fixed linking length FoF for all multiplicities. The MCL group catalogue has better halo completeness and group purity than the comparable FoF catalogues, with a completeness and purity that is approximately independent of multiplicity. As a result, MCL provides a better estimate of the number of groups of a given multiplicity than either of the two FoF models considered. In particular, compared to the best FoF approach (as measured by VI), it significantly improves the purity of, and the estimate of the number of, high multiplicity groups. This is most likely because MCL addresses better, through its inflation parameter, the problem of bridges linking large structures together, a common limitation of FoF. %(especially in Poisson noise limited samples).

MCL allows pairwise connection amplitudes that are not just ones and zeros, which may prove useful in catalogues with mixed redshift measurement precision, such as those from the PAU Survey~\citep[e.g.][]{eriksen2018}. Even in real space, where pairwise connections are not probabilistic, MCL produces better group catalogues than FoF. Future work will test MCL on more detailed mock galaxy catalogues in redshift space with photometric errors.

\section*{Acknowledgements}
We thank the referee, Elmo Tempel, for insightful comments. This work was supported by the Science and Technology Facilities Council 
[ST/J501013/1, ST/L00075X/1, ST/P000541/1]. 
PN acknowledges the receipt of a  Royal Society University Research Fellowship. We acknowledge support from the Royal Society international exchange programme.
This work used the DiRAC Data Centric system at Durham University,
operated by the Institute for Computational Cosmology on behalf of the STFC DiRAC HPC Facility \url{www.dirac.ac.uk}. This equipment was funded by
BIS National E-infrastructure cap- ital grant ST/K00042X/1, STFC
capital grant ST/H008519/1, and STFC DiRAC Operations grant
ST/K003267/1 and Durham University. DiRAC is part of the National E-Infrastructure.

%%%%%%%%%%%%%%%%%%%%%%%%%%%%%%%%%%%%%%%%%%%%%%%%%%

%%%%%%%%%%%%%%%%%%%% REFERENCES %%%%%%%%%%%%%%%%%%

% The best way to enter references is to use BibTeX:

\bibliographystyle{mnras}
\bibliography{references} % if your bibtex file is called example.bib

% Alternatively you could enter them by hand, like this:
% This method is tedious and prone to error if you have lots of references
%\begin{thebibliography}{99}
%\bibitem[\protect\citeauthoryear{Author}{2012}]{Author2012}
%Author A.~N., 2013, Journal of Improbable Astronomy, 1, 1
%\bibitem[\protect\citeauthoryear{Others}{2013}]{Others2013}
%Others S., 2012, Journal of Interesting Stuff, 17, 198
%\end{thebibliography}

%%%%%%%%%%%%%%%%%%%%%%%%%%%%%%%%%%%%%%%%%%%%%%%%%%

%%%%%%%%%%%%%%%%% APPENDICES %%%%%%%%%%%%%%%%%%%%%

%%%%%%%%%%%%%%%%%%%%%%%%%%%%%%%%%%%%%%%%%%%%%%%%%%

% Don't change these lines
\bsp	% typesetting comment
\label{lastpage}
\end{document}